# Evault for Legal Records

Abhishek S, Anas S, Anuragav R, Sachin K
*Department of Computer Science Engineering*
*Kumaraguru College of Technology,*
*Coimbatore – 641 049*
Corresponding author: jeba.n.cse@kct.ac.in

*Abstract*— Innovative solution for addressing the challenges in the legal records management system through a blockchain-based eVault platform. Our objective is to create a secure, transparent, and accessible ecosystem that caters to the needs of all stakeholders, including lawyers, judges, clients, and registrars. First and foremost, our solution is built on a robust blockchain platform like Ethereum harnessing the power of smart contracts to manage access, permissions, and transactions effectively. This ensures the utmost security and transparency in every interaction within the system. To make our eVault system user-friendly, we've developed intuitive interfaces for all stakeholders. Lawyers, judges, clients, and even registrars can effortlessly upload and retrieve legal documents, track changes, and share information within the platform. But that's not all; we've gone a step further by incorporating a document creation and saving feature within our app and website. This feature allows users to generate and securely store legal documents, streamlining the entire documentation process.

## I. INTRODUCTION

In the digital age, the legal industry is undergoing a profound transformation. Legal professionals, clients, and the judiciary increasingly rely on electronic documents and records to streamline operations and improve access to critical information. Yet, despite these advancements, challenges persist in ensuring the integrity, security, and accessibility of legal records. The need for a robust, tamper-proof, and transparent system for managing legal documents has never been more pressing.

The "eVault for Legal Records using Blockchain" project seeks to address these challenges by harnessing the power of blockchain technology to revolutionize the way legal records are created, stored, and accessed. Blockchain, the underlying technology behind cryptocurrencies like Bitcoin, offers a decentralized, secure, and immutable ledger, ideally suited to the sensitive and often critical nature of legal records.

This groundbreaking initiative aims to redefine the way legal records are managed, enhancing their security, transparency, and efficiency while ensuring compliance with legal and regulatory requirements. By creating a digital eVault for legal records on a blockchain, this project will pave the way for a future where legal professionals, clients, and authorities can rely on a cutting-edge system to safeguard the integrity of legal documents and streamline legal processes.

### A. BLOCKCHAIN TECHNOLOGY

Blockchain is a decentralized and distributed ledger technology that enables secure, transparent, and tamper-resistant record-keeping of transactions across a network of computers. It operates on a peer-to-peer network, where each participant (node) in the network has a copy of the entire blockchain. The information is stored in blocks, and each block is linked to the previous one through cryptographic hashes, creating a chain of blocks, hence the name "blockchain."

The key features of blockchain include:
*1)Decentralization:* No single entity or authority controls the entire network. Instead, participants collectively validate and agree on the state of the ledger.
*2)Transparency:* All participants have access to the same information, promoting openness and trust in the system.
*3)Immutability:* Once information is added to the blockchain, it is extremely difficult to alter or delete. This immutability is achieved through cryptographic hashing and consensus mechanisms.
*4)Security:* Blockchain uses advanced cryptographic techniques to secure transactions and ensure the integrity of the data.

### B. eVault for Legal System using Blockchain:

*1)Document Verification and Authenticity:* When a legal document is stored in the eVault, a unique cryptographic hash is generated for that document. This hash is like a digital fingerprint and is stored on the blockchain. Any modification to the document would change the hash, alerting all participants to the tampering attempt. This ensures that the document's authenticity and integrity can be verified by anyone with access to the blockchain, enhancing trust in the legal system.
*2)Smart Contracts*: In the legal system, smart contracts can automate and enforce the execution of agreements. For example, a real estate transaction could be facilitated through a smart contract that automatically transfers ownership once payment is confirmed. The self-executing nature of smart contracts reduces the need for intermediaries, streamlines processes, and minimizes the risk of disputes.
*3)Chain of Custody*: Blockchain's decentralized and transparent nature is particularly valuable in maintaining an immutable chain of custody for evidence or critical legal documents. Each transfer of custody is recorded as a transaction on the blockchain, creating a clear and traceable history. This feature is crucial in legal cases where the integrity and authenticity of evidence are paramount.
*4)Digital Identity and Access Control:* Blockchain can be used to establish a decentralized and secure system for managing digital identities. Each user could have a unique cryptographic key that serves as their digital identity. This not only reduces the risk of identity theft but also enables secure and granular access control to legal documents. Access permissions can be managed through smart contracts, ensuring that only authorized individuals can view or modify specific information.

*5) Time-Stamping and Notarization:* Blockchain's ability to timestamp transactions is valuable for notarization purposes in the legal system. Each transaction on the blockchain is associated with a specific timestamp, providing an immutable record of when an event occurred. This feature is crucial for establishing the chronology of events in legal proceedings, helping to resolve disputes and provide a reliable timeline.

*6) Consensus Mechanism:* The consensus mechanism in blockchain ensures that all participants agree on the state of the ledger. This agreement is reached through a process that involves validating transactions and adding them to the blockchain. Consensus mechanisms, such as Proof of Work or Proof of Stake, contribute to the security and trustworthiness of the system.

*7) Decentralization and Redundancy*: The decentralized nature of blockchain ensures that there is no single point of failure. Each participant in the network has a copy of the entire blockchain, creating redundancy. This makes the system resilient to cyber attacks or data loss, enhancing the overall reliability of the eVault for the legal system.

By leveraging these features, blockchain technology offers a robust and secure foundation for managing legal documents, automating processes, and enhancing the overall efficiency and trustworthiness of the legal system through an eVault.

*C. Django Framework:*

A table heading (using the "table head" style) appears above a table. This will automatically number the table for you. Any footnotes appear below the table, using the "table footnote" style. Footnotes are indicated by superscript lowercase letters within the table. An example of a table can be seen in Table I, below. A web application is a software application that runs on a web server, accessible through a web browser, and often designed to perform specific functions or provide certain services to users. One popular framework for building web applications in Python is Django. Django is a high-level, open-source web framework that encourages rapid development and clean, pragmatic design.

Django simplifies the process of building robust and scalable web applications by providing a set of tools, libraries, and conventions that streamline common development tasks. It follows the Model-View-Controller (MVC) architectural pattern, emphasizing the separation of concerns and promoting code reusability.

Here's a brief overview of the key components and concepts within a web app using the Django framework:

*1) Model:* Django uses an Object-Relational Mapping (ORM) system to define models, which represent the application's data structure. Models are Python classes that define the fields and behaviors of the data entities, making it easy to interact with databases without writing raw SQL queries.

*2) View*: Views in Django handle the presentation logic of the application. They receive requests from the user, process the data using the models, and return the appropriate response. Views can render HTML templates, handle form submissions, and interact with the underlying data.

*3) Template:* Django templates are used to define the structure and layout of the HTML pages presented to the user. Templates are powerful and allow for the dynamic rendering of data within HTML, creating a clean separation between the presentation layer and business logic.

*4) Controller (URLconf):* URLconf (URL configuration) maps URLs to the corresponding views in Django. It acts as a controller in the MVC pattern, directing incoming requests to the appropriate view function based on the defined URL patterns. This helps in organizing the application's URL structure in a readable and maintainable way.

*5) Middleware:* Middleware in Django allows for the processing of requests and responses globally before reaching the view or after leaving the view. It can be used for tasks such as authentication, security, and request/response modification.

*6) Admin Interface:* Django provides a built-in admin interface that automatically generates an administration panel for managing models and data. This feature makes it easy for developers to perform CRUD (Create, Read, Update, Delete) operations on the application's data without building a separate admin section.

*7) Forms and Authentication:* Django simplifies the creation of forms for user input and provides robust authentication mechanisms out of the box. Forms are used for data validation and handling user-submitted information, while authentication ensures secure user access to different parts of the application.

*8) Django ORM:* The Django ORM simplifies database interactions by abstracting the underlying database system. Developers can define models and perform database operations using Python code, and Django takes care of translating these operations into SQL queries.

By combining these elements, Django empowers developers to create feature-rich and maintainable web applications efficiently. Its emphasis on best practices, scalability, and security makes it a popular choice for building a wide range of web applications, from small projects to large-scale, enterprise-level systems.

## II. LITERATURE REVIEW

[1] Verma and Ashwin introduced NyaYa, a blockchain-based Electronic Law (EL) management system for digitized judicial investigations. NyaYa has four phases: stakeholder registration, case registration on a public blockchain with meta-hash keys, updates among law enforcement agencies, and case settlement via smart contracts. Simulation shows NyaYa outperforms traditional EL storage in various aspects like mining cost, query time, and trust probability, making it an effective solution for secure and efficient digital evidence management in the legal system.

[2] Victoria L. Lemieux paper is about Blockchain, a distributed ledger technology, aims to create immutable records of transactions. It's rapidly transforming various sectors like healthcare, real estate, and finance, promising

trustworthy and secure recordkeeping. The technology involves blocks of transaction records cryptographically chained together, ensuring transparency and detectability of alterations. Transactions involve transferring blockchain representations (tokens) of assets between addresses using public-private key pairs. Blockchains can be decentralized or centralized, and they can be public or permissioned. In recordkeeping, blockchain proves advantageous by detecting alterations and enhancing privacy through individual data control. However, challenges like scalability and legal concerns persist

[3]Maisha Afrida Tasnim, Abdullah Al Omar , Mohammad Shahriar Rahman and Md Zakirul Alam Bhuiyan introduces a system utilizing blockchain for secure storage and management of criminal records, emphasizing authenticity, data integrity, and protection against tampering. By integrating criminal records into a blockchain and employing a peer-to-peer cloud network for decentralization, the system aims to prevent unauthorized alterations and enhance overall data security. The proposed system allows authorized users like law enforcement and general users to efficiently manage and access criminal records. The decentralized data management process involves pre-registered users, digital signatures, encryption, and blockchain technology to ensure data authenticity and prevent tampering.

[4]Ali, Saqib, Wang Guojun, White Bebo, Cottrell and Roger Leslie proposed a blockchain-based data storage and access framework for PingER, a global Internet performance measurement project. Utilizing a permissioned blockchain and Distributed Hash Tables (DHT), metadata is stored on the blockchain while actual files are stored off-chain via DHT in a decentralized manner. This design reduces reliance on a centralized repository, offering decentralized storage, distributed processing, and efficient file retrieval. Permissioned blockchains enhance security by allowing only authenticated and authorized participants, making it suitable for various applications beyond cryptocurrencies, including peer-to-peer cloud storage. The framework aims to improve PingER's data management and accessibility.

[5]Olumide Malomo, Danda Rawat and Moses Garuba address the critical issue of offsite data recovery and security in the face of cyber-attacks and disasters. Cyber threats, especially related to data storage, have become highly sophisticated and challenging to defend against. Ransomware attacks targeting sensitive data have shown a significant rise, highlighting the need for secure offsite storage solutions. The paper proposes a Blockchain-enabled federated cloud computing framework that offers efficient, private, and secure offsite storage for digital assets. By leveraging blockchain technology and implementing strict access controls, the framework aims to enhance data security, prevent breaches, and outperform traditional approaches in terms of efficiency and effectiveness.

[6]Mark W. Storer, Kevin Greenan, Darrell D. E. Long and Ethan L. Miller address the challenge of balancing data security and storage efficiency in archival systems. Traditional deduplication, aimed at optimizing storage by eliminating redundant data, conflicts with encryption, which seeks to conceal data patterns. The paper proposes a solution by generating encryption keys consistently from data chunks, allowing deduplication while maintaining encryption. This ensures identical chunks encrypt to the same ciphertext and keys remain confidential. The approach supports both single-server and distributed storage systems, providing enhanced data security and efficient storage. The paper outlines the security mechanisms and evaluates the system's effectiveness in achieving secure deduplication.

[7]Danielle Batista, Ana Lara Mangeth , Isabella Frajhof, Paulo Henrique Alves, Rafael Nasser, Gustavo Robichez, Gil Marcio Silva and Fernando Pellon de Miranda explores the application of blockchain technology in maintaining and controlling the chain of custody in forensics, particularly focusing on physical evidence. A systematic literature review (SLR) was conducted, analyzing 26 resources discussing blockchain-based solutions for evidence chain of custody issues. The results revealed a lack of studies addressing the application of blockchain to solve problems related to the physical evidence chain of custody. The study emphasizes the need for future research to focus on addressing this gap. Keywords include chain of custody, blockchain, smart contracts, physical evidence, and forensics.

[8]Kamshad Mohsin discusses blockchain technology and its interaction with legal frameworks, particularly focusing on its implications for contracts, intellectual property, personal data protection, and global legalities. It highlights the dichotomy between enabling and prohibitive legislation concerning blockchain and explores its potential benefits and challenges. The emerging field of Blockchain Law is acknowledged, emphasizing the need to align technology with evolving legal requirements globally. The document also mentions the impact of blockchain on distributed ledger technology, urging compliance with diverse data regulations worldwide. Additionally, regulatory sandboxes and their role in fostering blockchain innovation are discussed.

[9]Victoria Lemieux introduces a novel framework for assessing the potential of blockchain and distributed ledger technology in delivering trustworthy and immutable recordkeeping, focusing on archival science principles. Blockchain's ability to create trusted, unalterable records without reliance on a central authority is drawing global interest, especially in applications like land transfers and healthcare records. The paper emphasizes the need to evaluate blockchain's capabilities through an archival science lens, which is crucial for use cases heavily reliant on secure and authentic recordkeeping, highlighting its relevance in assessing blockchain-based recordkeeping systems.

III. OBJECTIVES OF PROJECT

The objective of a project to create an eVault for legal records using blockchain is to leverage blockchain technology to enhance the management, security, and accessibility of legal records. The primary objectives of such a project may include:

*Security and Immutability*: Ensure the highest level of security and immutability for legal records, making it nearly impossible to alter or tamper with stored records. This helps maintain the integrity of legal documents.

*Transparency and Trust*: Establish a transparent and auditable record-keeping system that fosters trust among all stakeholders, including legal professionals, clients, and authorities.

*Efficiency and Accessibility*: Improve the efficiency of legal record management by enabling quick and secure access to records from anywhere, reducing administrative overhead, and streamlining legal processes.

*Reduced Costs*: Potentially lower the costs associated with physical record storage, retrieval, and manual document handling by transitioning to a digital, blockchain-based system.

*Legal Compliance*: Ensure compliance with legal and regulatory requirements related to record-keeping, data security, and privacy, thereby reducing the risk of legal disputes and penalties.

*Data Integrity*: Guarantee the long-term integrity of legal records, reducing the risk of loss or deterioration compared to traditional paper-based records.

*Collaboration and Interoperability*: Promote collaboration and interoperability among different stakeholders in the legal ecosystem, such as law firms, courts, notaries, and clients.

*User-Friendly Interface*: Provide a user-friendly interface for legal professionals and clients to access, manage, and contribute to legal records, making the system accessible to a wide range of users.

*Reduced Fraud:* Mitigate the risk of document fraud, forgeries, and disputes by implementing strong cryptographic and authentication measures.

*Preservation of Legal History*: Ensure that historical legal records are preserved in a secure and reliable manner for reference, research, and legal precedent.

*Scalability and Longevity:* Create a platform that can scale to accommodate a growing volume of records and adapt to evolving blockchain technology to ensure the system's longevity.

## IV. EXISTING SYSTEM

### A. EXISTING METHOD

Legal records pertaining to court cases are traditionally stored physically in the respective courts where the cases were heard. However, some courts have adopted Electronic Case Management Systems (CMS) to digitize and manage court records. These systems often convert paper-based records into electronic formats, commonly in the form of PDF documents. Additionally, the National Judicial Data Grid (NJDG) platform is utilized in some regions for storing legal records, specifically focusing on case details.
Limitations:

*1) Limited Accessibility and Risk of Damage*: The physical storage of legal records restricts accessibility, making it challenging for authorized individuals to retrieve information quickly. Moreover, there is a risk of damage or loss associated with physical records, which could impact the integrity of the legal documentation.

*2) Compatibility Issues:* Different courts may use varied CMS software and systems, leading to compatibility issues when attempting to share information. This lack of standardization can hinder seamless collaboration and data exchange between different judicial entities.

*3) NJDG Limitations*: While the NJDG platform is in use, it focuses primarily on storing case details and may not encompass all relevant information, especially evidence. This limitation poses a challenge in maintaining a comprehensive and centralized repository of legal records.

*4) Chain of Custody Concerns*: In the existing system, there's a risk of evidence tampering during the chain of custody. The lack of a secure and transparent mechanism for tracking the movement and handling of evidence could compromise its integrity.

*5) Digitalization Discrepancies:* The process of digitalizing legal records varies from one court to another, leading to inconsistencies in the quality and completeness of the digital data. This lack of uniformity can hinder the effectiveness of a centralized digital records system.

*6) Multiple Signatures:* Clients may be required to sign legal statements multiple times, contributing to redundancy and potential errors in the documentation process. This manual approach can be time-consuming and may result in inefficiencies.

*7) Specialized Courts Integration:* Certain specialized courts, such as family courts, consumer forums, and labor tribunals, are not fully integrated into the NJDG. This lack of integration can create silos, making it difficult to have a holistic view of legal proceedings across different court types.

*8) Manual Legal Statement Writing:* The manual process of writing legal statements is time-consuming and may lead to delays in legal proceedings. Automation and digitalization of this process could significantly improve efficiency and reduce the likelihood of errors.

The existing method faces challenges related to limited accessibility, compatibility issues, incomplete digitalization, evidence integrity concerns, and lack of integration for specialized courts. Addressing these limitations would require the adoption of more standardized and technology-driven solutions to enhance efficiency, transparency, and overall effectiveness in managing legal records.

## V. PROPOSED SYSTEM

### A. PROBLEM IDENTIFICATION

To develop a blockchain-based eVault system for legal records that can ensure security, transparency, and accessibility for all stakeholders and integrate it with existing legal databases and case management systems.

## B. BLOCK DIAGRAM

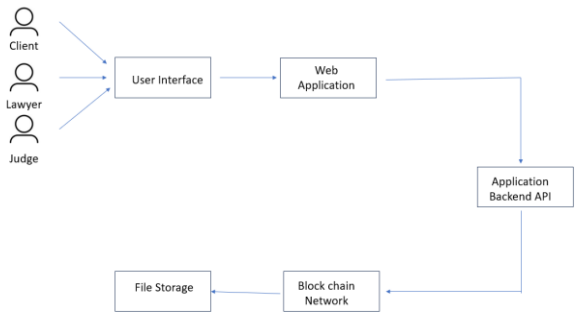

Client/Lawyer/Judge: These are the end-users or legal professionals who interact with the system. They can be individuals or entities involved in legal processes, such as clients seeking legal services, lawyers providing legal assistance, and judges overseeing legal cases.

User Interface: This component is the graphical or user-friendly interface that the client, lawyer, or judge uses to interact with the system. It provides a platform for users to input data, request information, and perform various actions related to legal matters.

Web Application: The web application is the software that runs on a web server and provides the user interface. It processes user requests, handles user authentication, and communicates with the application backend API to perform various functions related to legal services, such as document management, case tracking, communication, and more
.
Application Backend API: This is the part of the system responsible for processing requests from the web application and interacting with other components. It performs various functions, such as retrieving and storing data, executing business logic, and interfacing with the blockchain network for data storage.

Blockchain Network: The blockchain network is a decentralized and distributed ledger technology. It is used for secure and tamper-resistant data storage. In this architecture, it is used for storing legal documents, contracts, or other sensitive information. Blockchain technology ensures the integrity and immutability of data, making it suitable for legal applications where data security and authenticity are crucial.

File Storage: This component represents a storage system, integrated with the blockchain network. It's used for the long-term storage of legal documents, case records, and other files. Files stored in the blockchain are cryptographically secured and can be easily retrieved when needed.

## C. MODULES

An eVault system for legal records using blockchain typically consists of four key modules to ensure the security, integrity, and accessibility of legal records.

*1)Client Module:* The client module is the interface through which users, such as law firms, legal professionals, or clients, interact with the eVault system. It includes software applications or web interfaces that allow users to upload, access, and manage legal records securely.
Users can initiate actions like uploading legal documents, searching for records, granting access permissions, and viewing the history of document changes.
The client module may provide encryption and authentication mechanisms to ensure that only authorized users can access and modify legal records.

*2)Server Module:* The server module is responsible for managing the centralized or distributed server infrastructure that hosts the blockchain network and the eVault system's databases. It handles data storage, replication, and synchronization across the blockchain nodes, ensuring that legal records are stored securely and redundantly.
The server module also includes components for monitoring the health and performance of the system, managing system resources, and handling user requests.

*3)Docker Module:* Docker is a containerization technology that helps manage and deploy applications and their dependencies in isolated containers. In the context of an eVault system for legal records, the Docker module may serve several purposes:It allows for the containerization of various components of the system, making it easier to manage and scale the system. Docker containers can be used to isolate specific services or functions, such as document processing, data validation, or record verification.
Docker containers can provide an additional layer of security by isolating different parts of the system.

*4)Smart Contract Module*: Smart contracts are self-executing contracts with the terms of the agreement directly written into code. In the context of a legal eVault system, smart contracts play a crucial role in automating and enforcing the rules and policies related to record storage and access.The smart contract module consists of the smart contracts themselves, which are deployed on the blockchain. These smart contracts define the rules for creating, updating, and accessing legal records. They may also include rules for access control and record validation.
Smart contracts ensure transparency, security, and trust in the system by automating processes like verifying document authenticity and controlling access permissions.
Users interact with the smart contracts to create, modify, or query legal records, and the blockchain network enforces the contract terms.

## D. METHODOLOGY
 *1) Judge*

*a) Login:* The judge logs in using their Unique ID.

*b) Homepage:* Upon successful login, the judge is directed to a homepage displaying all pending cases in their jurisdiction. Cases are sorted according to the date of the next hearing, ensuring an organized view. Each case is listed with essential details such as case number, parties involved, and the next hearing date.

*c) Case Details:* By selecting a specific case, the judge can access detailed information about that case, including case specifics and any supporting documents. The system allows the judge to make notes, record decisions, or update the case status.

*2) Lawyer:*

*a) Login:* The lawyer logs in using their Unique ID.

*b) Homepage:* The lawyer's homepage displays all upcoming cases they are involved in. The cases are listed with relevant information such as case number, parties, and upcoming court dates.

*c) Upload Functionality:* Lawyers have the ability to upload necessary paperwork, evidence, or any other required documents for each case directly from their homepage. This feature streamlines the process of submitting and managing case-related documents.

*3) Ordinary Citizen:*

*a) Login:* The ordinary citizen logs in using their Unique ID.

*b) Homepage:* The homepage displays case details linked to the citizen's Aadhar number.

If there are ongoing cases, the system provides information such as case number, status, and relevant updates.

*c) Search Functionality:* An interactive search bar allows citizens to look for specific cases. Citizens can read details about cases, access court documents if public, and stay informed about legal proceedings. This feature empowers citizens to actively engage with the legal process.

Upon user signup, the system triggers an identity management function, initiating the creation of a digital identity for the user. The details obtained from the user's real identity are encrypted using the SHA-3 (Keccak-256) algorithm, ensuring the security and privacy of the user's information. This step is crucial for establishing a secure and trustworthy environment. After the successful authentication of the user, they are redirected to the homepage, where a centralized hub of past and current case details associated with the user is presented. Each case is represented as a block, and users can click on these blocks to access detailed information about their cases. This user-friendly interface enhances accessibility and transparency for individuals engaging with the legal system.

An additional feature is provided for authorized officials, allowing them to upload case documents directly through an upload button. This functionality reduces the administrative burden on lawyers and ensures that relevant case documents are efficiently integrated into the system. These documents are stored securely on the blockchain network, ensuring transparency, tamper-proofing, and increased data integrity.

To facilitate accessibility for normal citizens seeking legal assistance, a search button is incorporated on the homepage. Users can search for lawyers based on their specific situation and location, streamlining the process of finding legal professionals. Simultaneously, officials can leverage a machine learning algorithm integrated into the search function to obtain relevant case citations, optimizing their workflow and decision-making processes.

Smart contracts play a pivotal role in reducing the workload of lawyers. These contracts are employed for document management, enabling clients to use digital signatures for signing legal documents remotely. This feature is particularly advantageous for clients who may face challenges related to education or language barriers, as it allows them to sign documents from the comfort of their own space. To further enhance accessibility, a Text-to-Speech module is incorporated, benefiting clients with poor education levels or language barriers. This feature ensures that information is communicated effectively, regardless of the user's literacy or language proficiency.

## VI. CONCLUSION

The project introduces a robust and secure platform tailored for legal professionals, encompassing clients, lawyers, and judges, with a focus on streamlining interactions and managing legal processes. The innovative integration of blockchain technology for file storage elevates the system's capabilities by ensuring the integrity and immutability of critical legal documents and data. This blockchain-backed infrastructure safeguards against unauthorized alterations or tampering, offering a tamper-proof repository for legal records. The user-friendly web application enhances the overall user experience, promoting efficient communication and seamless information retrieval. Through a well-designed backend API, the system facilitates smooth interactions, real-time updates, and collaboration among legal professionals, enhancing the platform's scalability and adaptability. With an emphasis on data security, transparency, and reliability, this comprehensive solution stands as a promising advancement for the legal industry, facilitating the administration of legal cases and services while ensuring a trustworthy and efficient workflow for all stakeholders.

## VII. RESULTS

Implementing an eVault for legal records using blockchain technology offers numerous advantages. Blockchain's decentralized and immutable ledger ensures the integrity of records, safeguarding them against tampering or unauthorized access. The cryptographic techniques employed by blockchain enhance security, maintaining the confidentiality of sensitive legal data. Transparency and traceability are inherent features, providing all stakeholders with the ability to verify the authenticity of records and track their history. By digitizing record-keeping processes and automating tasks through smart contracts, organizations can realize efficiency gains, reduce administrative overheads, and streamline operations. Furthermore, blockchain-based eVaults facilitate compliance with regulatory requirements by offering a secure and auditable solution for record management. The global accessibility of blockchain enables seamless access to legal records from anywhere, promoting collaboration and efficiency across geographically dispersed teams. While the benefits are promising, successful implementation necessitates careful consideration of factors such as regulatory compliance, data privacy, and interoperability. Nonetheless, the potential of blockchain-based eVaults to revolutionize legal record management is significant, offering a robust and future-proof solution for organizations seeking to modernize their processes.

A. SCREENSHOTS

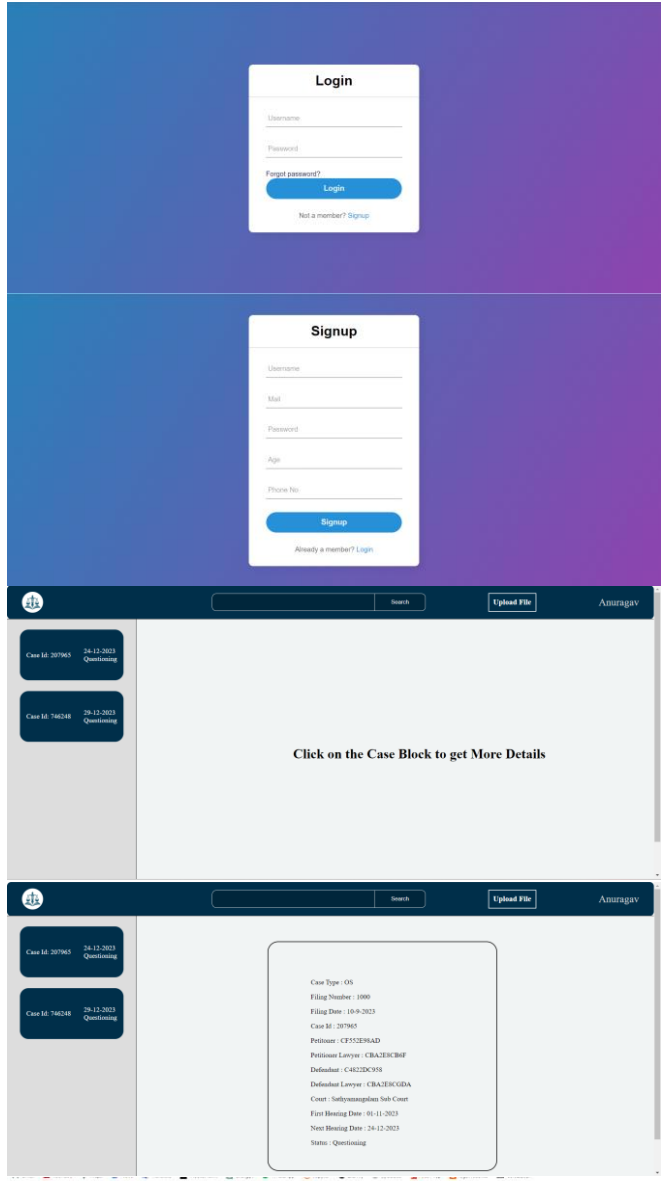

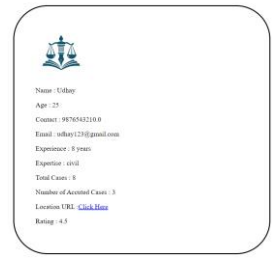

## REFERENCES

During the preparation of this work the author(s) used chatgpt in order to nhance natural language understanding and generate conversational content for analysis. After using this tool/service, the author(s) reviewed and edited the content as needed and take(s) full responsibility for the content of the publication